\documentclass[prl,aps,twocolumn,superscriptaddress,reprint]{revtex4-1}
\pdfoutput=1
\usepackage{color}
\usepackage{graphicx}
\usepackage{amsmath}
\usepackage{amssymb}
\usepackage{cancel}
\usepackage{diagbox}
\usepackage{caption}
\graphicspath{{./figs/}}
\makeatletter
\def\input@path{{./figs/}}
\makeatother

\newcommand{\comment}[1]{}
\newcommand{\ig}[2]{\includegraphics[width = #1]{#2}}

\newcommand{\wpwc}{\omega_{pe}/\omega_{ce}}

\begin{document}
\title{Kinetic simulations underestimate the effects of waves during magnetic reconnection}
\author{J. Ng}
\affiliation{Department of Astronomy, University of Maryland, College Park, MD 20742, USA}
\affiliation{NASA Goddard Space Flight Center, Greenbelt, MD 20771}
\author{J. Yoo}
\affiliation{Princeton Plasma Physics Laboratory, Princeton, NJ 08540}
\author{L.-J. Chen}
\affiliation{NASA Goddard Space Flight Center, Greenbelt, MD 20771}
\author{N. Bessho}
\affiliation{Department of Astronomy, University of Maryland, College Park, MD 20742, USA}
\affiliation{NASA Goddard Space Flight Center, Greenbelt, MD 20771}
\author{H. Ji}
\affiliation{Princeton Plasma Physics Laboratory, Princeton, NJ 08540}

\date{\today}
\begin{abstract}
  Collisionless plasma systems are often studied using fully kinetic simulations, where protons and electrons are treated as particles. Due to their computational expense, it is necessary to reduce the ion-to-electron mass ratio $m_i/m_e$ or the ratio between plasma and cyclotron frequencies in simulations of large systems. In this work we show that when electron-scale waves are present in larger-scale systems, numerical parameters affect their amplitudes and effects on the larger system. Using lower-hybrid drift waves during magnetic reconnection as an example, we find that the ratio between the wave electric field and the reconnection electric field scales like $\sqrt{m_i/m_e}$, while the phase relationship is also affected. The combination of these effects means that the anomalous drag that contributes to momentum balance in the reconnection region can be underestimated by an order of magnitude. The results are relevant to the coupling of electron-scale waves to ion-scale reconnection regions, and other systems such as collisionless shocks. 
\end{abstract}

\maketitle

In collisionless plasma systems, such as those found in space and astrophysical environments, electron and ion distribution functions show quite some deviation from the Maxwell-Boltzmann distribution (e.g.~\cite{graham:2021,shi:2022,burch:2016,raptis:2022,Maksimovic:2005,schekochihin:2005}). The use of the Vlasov equation, with three dimensions in velocity space, and three in position space, is necessary to understand the complex dynamics of such physical systems. Kinetic simulations that evolve the particle distributions or approximate the distributions using macroparticles (e.g.~\cite{bowers:2008,markidis:2010,fonseca:2002,fujimoto:2018,juno:2018}) are a powerful tool used to solve the Vlasov equation.

In spite of the advances in modern computing capabilities, fully kinetic simulations, which treat ions and electrons as particles, are still limited. Many physical systems require multiple spatial and temporal scales to be resolved, leading to compromises in the physical parameters used in simulations. Using Earth's magnetosphere as an example, studies of localised magnetic reconnection regions involve dimensions of tens to hundreds of ion inertial lengths in two- or three-dimensions (e.g.~\cite{egedal:2012,le:2017,price:2020,lapenta:2018} ), while studies of the collisionless shock can go from hundreds to up to over a thousand ion inertial lengths (e.g.~\cite{lu:2024,savoini:2015,bessho:2022}). Other simulations studying kinetic phenomena in the magnetotail use an artificially smaller system, yet still require scales of tens to hundreds of ion inertial lengths \cite{totorica:2023,liu:2014}. To simulate these scales while still resolving electron kinetic physics, a reduced electron-ion mass ratio is generally employed, reducing the separation of scales, and the ratio between electron plasma and cyclotron frequencies is also reduced, reducing the ratio between the speed of light and the electron Alfv\'en speed $c/V_{Ae}$.

Kinetic simulations have been successful in the study of magnetic reconnection, and the use of reduced parameters in such simulations is supported by the result that during collisionless reconnection at ion spatial scales, the reconnection rate is insensitive to the exact electron physics \cite{comisso:2016, birn:2001, cassak:2017:01}. However, there are instances where the numerical parameters have affected the results qualitatively. It was shown that large mass ratios allowed the development of a new regime of reconnection with embedded exhaust current layers \cite{le:2013}, while Ref.~\cite{jaraalmonte:2014} showed that increasing the frequency ratio leads to Debye scale turbulence developing in the reconnecting layer. Artificially low mass ratios also allow the development of drift-kink instabilities that disrupt current sheets \cite{daughton:1998}. The use of reduced parameters also significantly affects results in other fields such as shock physics. It is known that observations of electric field fluctuations in shock crossings are much stronger than those found in simulations because of the use of reduced parameters \cite{wilson:2021,bohdan:2024}.

In this work we use the specific example of lower-hybrid drift waves during magnetic reconnection to quantitatively study the effects of numerical parameters and their importance when coupling electron-scale waves to the ion-scale reconnection region and beyond. Lower-hybrid drift waves are driven by the diamagnetic drift \cite{davidson:1977,huba:1976,krall:1971,davidson:1975} and are often found in magnetic reconnection regions. Their importance in the reconnection region is an area of active research \cite{graham:2022,mozer:2011,yoo:2024,chen:2020}, where they can contribute to electron momentum balance through correlated density and electric field fluctuations known as anomalous drag, or cause electron heating. 


There are numerous simulation, experimental and observational studies of the role of the lower-hybrid drift instability (LHDI) during magnetic reconnection, in which it is shown that they cause particle transport and mixing (e.g.~\cite{le:2017,le:2018,price:2020}). Although it is known that generating the waves requires proper scale separation between electrons and ions \cite{daughton:2003, roytershteyn:2012, price:2016}, existing studies do not discuss saturation and how numerical parameters affect calculations of the waves' contribution to momentum balance in the larger reconnection region \cite{le:2017,le:2018,price:2016,roytershteyn:2012}. Theoretical studies of the waves themselves \cite{winske:1978} consider their saturation, but do not account for their coupling to magnetic reconnection. These limitations mean that quantitative comparisons between simulations and observations or experiments are not adequate when considering the broader effects of the waves. 



Recently, it has been shown in experiments and observations that during guide-field reconnection (where there is a magnetic field parallel to the initial current sheet), the LHDI is excited by the electron flow in the reconnection exhaust \cite{chen:2020,yoo:2024}. While waves are seen in simulations of these events, there are still discrepancies in both particle acceleration and electron momentum balance related to the wave amplitudes \cite{ng:2023,ng:2020}. We perform simulations using the guide-field reconnection configuration to study the effects of numerical parameters on wave amplitudes and momentum balance. Due to computational limitations, we then use an alternative setup to study the waves driven by the LHDI in isolation and evaluate their scaling with numerical parameters systematically. Our results show that the normalized wave amplitudes and anomalous drag increase as parameters become more realistic.

We first give a brief introduction to the theory of LHDI saturation and how it relates to the reconnection electric field. The LHDI is driven by a diamagnetic current drifting across the magnetic field \cite{davidson:1977,huba:1976,krall:1971}. The simplest estimate of the saturation amplitude of the LHDI is given in \cite{winske:1978} as
\begin{equation}
  \mathcal{E} = n m_e V_d^2/\left[4\left(1 + \omega_{pe}^2/\omega_{ce}^2\right)\right],
  \label{eq:saturation_real}
\end{equation}
where $\mathcal{E}$ is the electric field energy density, $n$ is the local density, $V_d$ is the relative velocity between electrons and ions, $m_e$ is the electron mass, and $\wpwc$ is the ratio between electron plasma and cyclotron frequencies. This equates the free energy to the wave and particle energy in the electrostatic limit. Other saturation mechanisms such as ion trapping exist and will be discussed later \cite{brackbill:1984,winske:1978,lavorenti:2021}.

To compare the wave fields to magnetic reconnection, we note that in the collisionless limit, the reconnection electric field is generally around $0.1 B_0 v_{A0}$, where $B_0$ is the upstream magnetic field and $v_{A0}$ is the ion Alfv\'en speed calculated using upstream quantities \cite{comisso:2016, birn:2001, cassak:2017:01}. In the guide-field configuration we study, $V_d$ is given by the difference between the electron and ion outflow speeds, and is proportional to the electron Alfv\'en speed $v_{Ae0} = B_0/\sqrt{\mu_0 n_0 m_e }$, where $n_0$ is the upstream electron density \cite{karimabadi:2007}. This can be substituted into the expression for the energy density, and after some manipulation, we find
\begin{equation}
  \frac{\delta E}{B_0 v_{A0}} \sim \frac{\omega_{ce}}{\omega_{ce0}} \frac{\omega_{pe}}{\omega_{ce}\sqrt{1 + (\wpwc)^2}}\sqrt{\frac{m_i}{m_e}}.
  \label{eq:saturation}
\end{equation}
The first term $\omega_{ce}/\omega_{ce0}$ is the ratio of the local to upstream electron cyclotron frequency and can be treated as a scaling factor in this study, and other quantities on the right-hand side are measured locally. This expression explicitly scales with the mass ratio and the frequency ratio while the normalized reconnection electric field is not sensitive to the mass and frequency ratios.

We perform two types of simulations to study the variation of the lower-hybrid waves with numerical parameters -- ``reconnection'' simulations and ``wave'' simulations. The two-dimensional reconnection simulations use the same physical parameters as \cite{ng:2023} where the initial conditions consist of a Harris sheet superposed on an asymmetric background, while wave simulations capture the growth of lower-hybrid waves in a current layer. The reconnection simulations illustrate the coupling of lower-hybrid waves to the reconnection process, while the wave simulations allow a study using more realistic parameters.

We first study the development of lower-hybrid waves during asymmetric guide-field reconnection, the initial conditions for the base case are exactly the same as \cite{ng:2023}, which was based on MRX experiments. The asymmetric layer is defined by $B_L/B_H = 1.25$, $T_{eL} = T_{eH} = T_{iH}$, $n_L/n_H = 0.5$ and $T_{iL}/T_{eH} = 1.23$ where the subscripts represent the asymptotic values on either side of the current sheet. Were this magnetopause reconnection, $L$ would correspond to the magnetosphere and $H$ the magnetosheath. A constant guide field $B_g = 1.8 B_H$ is present, and the electron beta on the high density side is $\beta_{eH} = 2 \mu_0 n_H T_{eH}/(B_H^2 + B_g^2) = 0.3$. Instead of performing 3D simulations, we rotate the $x$-$z$ plane in our two-dimensional simulations by $5.7^\circ$ so that the magnetic field is almost perpendicular to the simulation plane in the unstable region found in \cite{ng:2023}, which is favorable for the excitation of lower-hybrid waves.

\begin{figure}
\ig{3.375in}{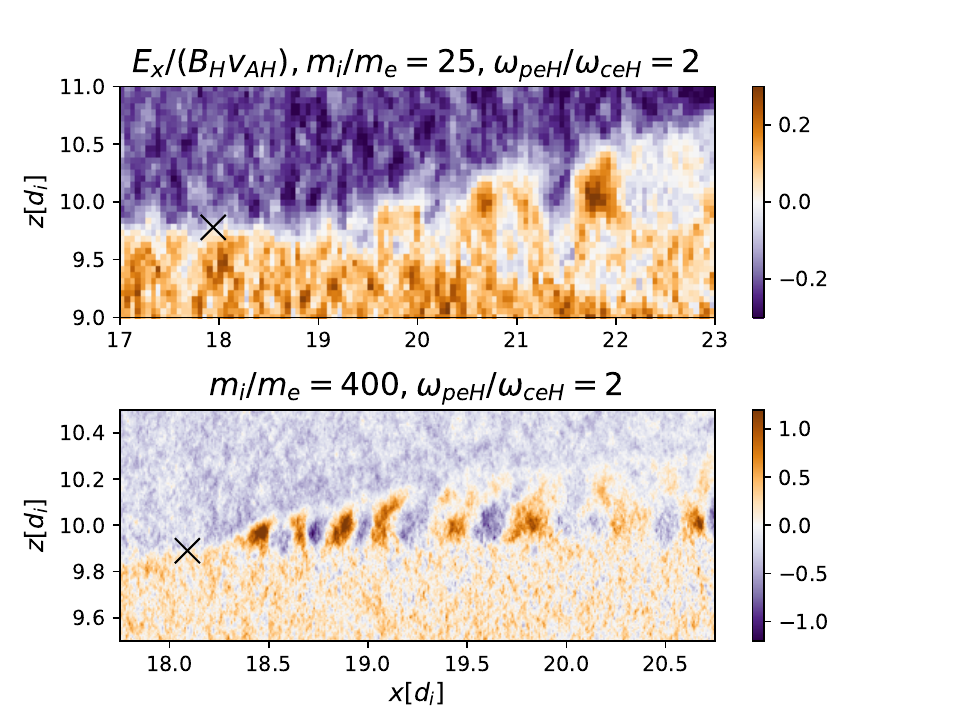}
\caption{Normalized wave electric field $E_x/(B_0v_{A0})$ in the $m_i/m_e = 25$, $\omega_{pe0}/\omega_{ce0}=2$ and $m_i/m_e = 400$, $\omega_{pe0}/\omega_{ce0}=2$ simulations. `X' marks the locations of the x-lines.} 
\label{fig:ey_rec}
\end{figure}

We show the results from two simulations with $m_i/m_e = 25$ and $m_i/m_e = 400$, both having $\omega_{peH}/\omega_{ceH} = 2$ (the realistic value is approximately 125 in the MRX experiment). Both simulations have dimensions $L_x\times L_z = 40 d_i\times 20 d_i$, using $1536\times 768$ and $12288\times 6144$ grid cells respectively. The asymmetric conditions cause a density gradient across the outflow region, and the initial configuration has the plasma beta lowest in the upper-right quadrant. Lower-hybrid drift waves develop, similar to the original 3D simulation with $m_i/m_e = 100$ \cite{ng:2023}. The fluctuating electric fields of the lower-hybrid waves are illustrated in Figure~\ref{fig:ey_rec}. The normalized wave amplitudes are clearly higher in the mass ratio 400 simulation. The main reason for the larger normalized amplitude is because the normalized outflow $u_{ex}/v_{A0}$ increases as mass ratio increases as implied by Equation~\eqref{eq:saturation}.

Although these results suggest that the normalized wave electric field increases as realistic parameters are approached, it is challenging to perform a systematic scan because these  waves only appear after reconnection has developed, and there are differences in evolution between simulations (such as plasmoid formation that changes the overall structure of the reconnection region). As such, we perform a set of simulations focused only on the generation and evolution of the waves.

The initial setup for a wave simulation is a Vlasov confinement equilibrium with a narrow current layer in the center of the $x$ domain \cite{hewett:1976}. This has previously been used for studies of the contribution of the LHDI to anomalous resistivity \cite{winske:1978}. The ion density is given by 
\begin{equation}
n_i(x) = \frac{1 + \epsilon - \tanh\left(\alpha\left(\frac{x^2}{a^2} -1\right)\right)}{1 + \epsilon + \tanh\left(\alpha\right)},
\end{equation}
where
\begin{equation}
\epsilon = \frac{ n_\infty\left(1+\tanh\left(\alpha\right)\right)}{1 - n_\infty}.
\end{equation}
The ion temperature is uniform and isotropic, and the electron temperature components $T_{xx}$ and $T_{zz}$ are uniform, while the initial magnetic field, electric field and other electron quantities are determined self-consistently using the recursive relations derived in Ref.~\cite{hewett:1976}. We set $\alpha = 5$, $a = 20$ and $n_\infty = 0.4$, where $\alpha$ and $a$ control the thickness of the current layer and $n_\infty$ is the asymptotic density on the positive $x$ side of the simulation. The electron distributions are expressed as a sum of Hermite polynomials. We cut off the infinite sum at 4 as adding higher-order terms does not change the distribution significantly. Similar configurations have been used to study the LHDI in other contexts \cite{lavorenti:2021}. Note that the recursive relations here do not depend on the mass ratio. 

\begin{figure}
\ig{3.375in}{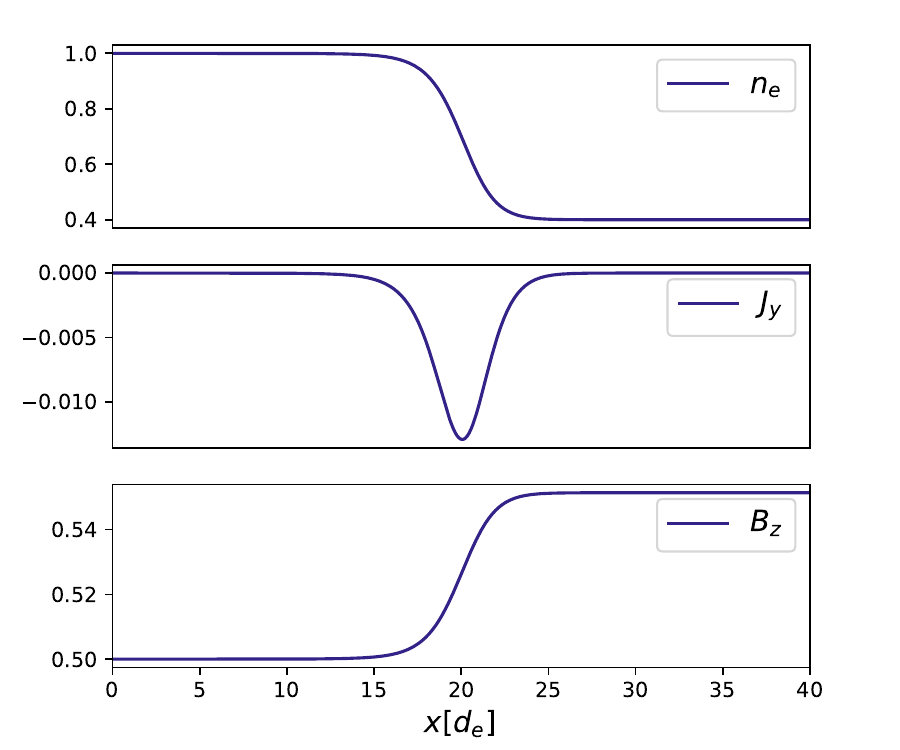}
\caption{Initial electron density, current and magnetic field profiles in wave simulations.}
\label{fig:initial_sheath}
\end{figure}

The computational domain is $L_x \times L_y = 40 d_e \times 50 d_e$ covered by $320 \times 450$ cells, and the plasma parameters for the baseline case defined at $x = 0$ are $\omega_{pe0}/\omega_{ce0} = 1$, $T_{i0}/T_{e0} = 1.25$ and $\beta_e = 0.16$.  The ion-to-electron mass ratio is $m_i/m_e = 100$ and there are 400 particles per species per cell. These parameters are chosen so the plasma in the current layer is similar to the unstable region in \cite{ng:2023}. These simulations are much smaller than the reconnection simulations, allowing us to perform a wider parameter scan by varying $m_i/m_e$ from 100 to 1836, and $\omega_{pe}/\omega_{ce}$ from $0.5$ to $8$. The resolution is increased appropriately as simulation parameters change so that the electron Debye length is resolved. The number of particles per cell is increased to 3200 for the $\omega_{pe0}/\omega_{ce0} = 8$ simulations. The initial conditions are shown in Figure~\ref{fig:initial_sheath} for $m_i/m_e = 1836$, $\omega_{pe0}/\omega_{ce0} = 2$.

\begin{figure}
\ig{3.375in}{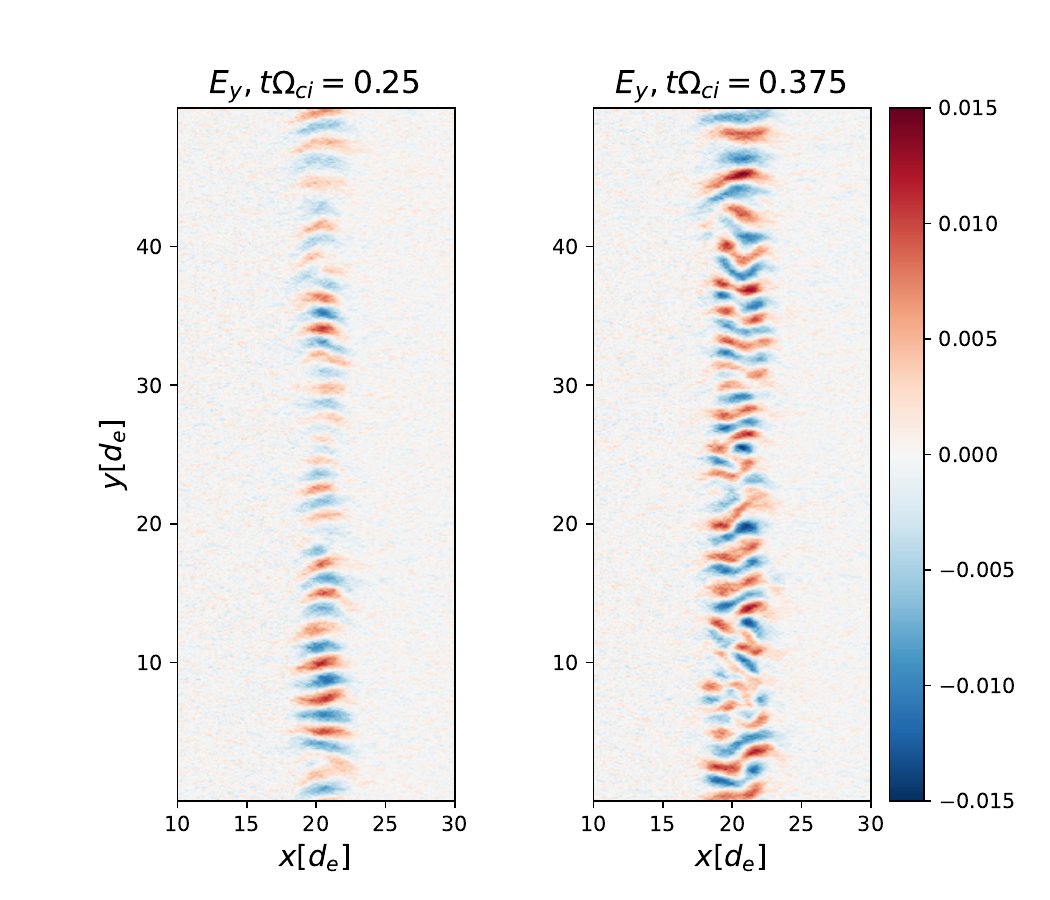}
\caption{Electric field $E_y$ in the $m_i/m_e = 1836$, $\omega_{pe0}/\omega_{ce0}=2$ simulation.}
\label{fig:sheath_ey}
\end{figure}

The evolution of the system is shown in Figure~\ref{fig:sheath_ey} for the simulation with $m_i/m_e = 1836$ and $\omega_{pe0}/\omega_{ce0}=2$. In this case, the lower-hybrid drift waves develop in less than one ion-cyclotron time, shown by the $E_y$ signatures in the center of the domain. The peak fluctuation amplitudes are measured as the current layer starts to break up, corresponding to $t\Omega_{ci} = 0.375$ in the Figure. This behavior is similar in all simulations, though the instability growth rate (is proportional to $\omega_{LH}$ \cite{davidson:1977}) relative to the ion cyclotron frequency increases with the mass ratio as $\omega_{LH}/\Omega_{ci} \propto \sqrt{m_i/m_e}$. 

For each simulation, we calculate the root mean square of the fluctuating electric field $\delta E_{y,rms}$, and the anomalous drag term $\langle\delta n_e \delta E_y\rangle/\langle n_e\rangle$ by integrating along the $y$ direction. The maximum values are then evaluated along $x$. For comparison between different simulations, these quantities are normalized by $B_0 v_{A0}$ evaluated at $x = 0, t=0$ in each simulation. 

\begin{figure}
\ig{3.375in}{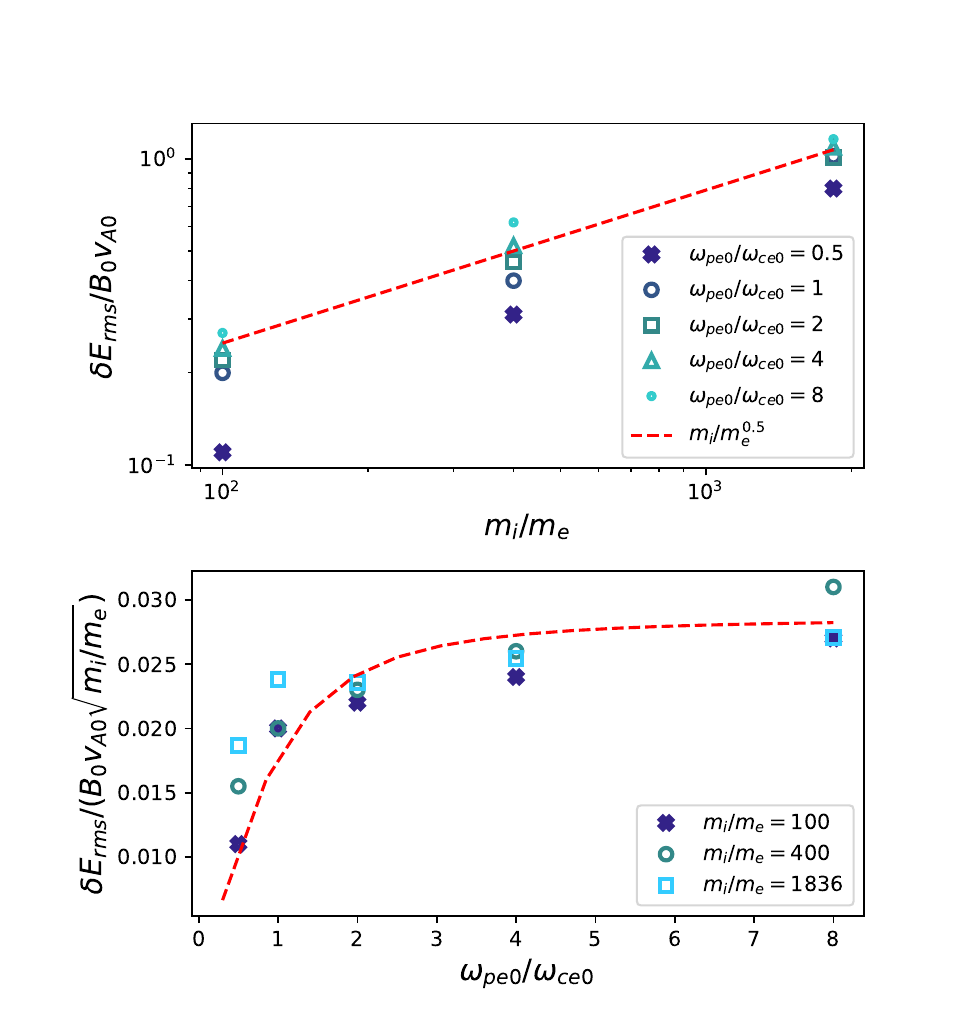}
\caption{Comparison between electric field fluctuation amplitude and scaling estimate. Top: Variation with mass ratio. Bottom: Variation with initial $\omega_{pe0}/\omega_{ce0}$ in simulations. The dashed line is proportional to the local $\omega_{pe}/\omega_{ce}/\sqrt{1+(\omega_{pe}/\omega_{ce})^2}$ where the plasma frequency is calculated using $n_e = 0.7$ and the cyclotron frequency is calculated using $B = 1.05 B_0$. } 
\label{fig:sheath_data}
\end{figure}

The maximum values of the normalized $\delta E_{y,rms}$ during the simulation are displayed in Figure~\ref{fig:sheath_data}. For the electric field fluctuations, there is a clear increase in the normalized values with mass ratio, in addition to a weaker variation with $\omega_{pe}/\omega_{ce}$.

Fitting the variation with mass ratio gives a scaling of $(m_i/m_e)^{0.56}$ and a correlation coefficient $r = 0.98$, showing good agreement with the $\sqrt{m_i/m_e}$ expression in Equation~\eqref{eq:saturation}, indicating the importance of the mass ratio to the normalized wave amplitudes in simulations.

The variation of the electric field amplitude does not show exact agreement with the $\wpwc/\sqrt{1+(\wpwc)^2}$ relationship, but the general trend of a steeper increase at small $\wpwc$ and smaller increase at larger, more realistic conditions holds. Equation~\eqref{eq:saturation} does not include electron temperature or electromagnetic effects, which partially account for the larger discrepancies at $\wpwc \leq 1$. 


The largest initial ratio between the drift velocity and ion thermal speed $V_d/v_{thi} \approx 3.5$ is found in the $m_i/m_e = 1836$ simulations, which is within the empirical limits suggested by Ref.~\cite{winske:1978} for the validity of Equation~\eqref{eq:saturation_real}, and close to the $V_d/v_{thi} \approx 3$ boundary between Equation~\eqref{eq:saturation_real} and the ion trapping saturation mechanism found by \cite{brackbill:1984}. While we see accelerated ions (not shown), there is surprisingly good agreement with Equation~\eqref{eq:saturation_real} at the larger $\wpwc$ relevant to laboratory and magnetospheric conditions. The electron resonance is unlikely to be important due to the low beta \cite{huba:1978}.  


\begin{table}[h]
  \caption*{$\langle \delta E_y \delta n_e\rangle/(\langle n_e \rangle B_0 v_{A0})$}  \begin{tabular*}{3.375in}{@{\extracolsep{\fill}} c c c c c c}
    \hline\hline
    \diagbox{$m_i/m_e$}{$\omega_{pe0}/\omega_{ce0}$} & 0.5 & 1 & 2 & 4 & 8 \\
    \hline
100 & 0.0033 & 0.0041 & 0.0033 & 0.0041& 0.0042\\
400 & 0.0090 & 0.0089 & 0.013 & 0.013 & 0.016 \\
1836 & 0.034 & 0.044 & 0.038 & 0.03 & 0.031 \\
\hline\hline
\end{tabular*}
\caption{Normalized amplitude of the anomalous drag term $\langle \delta E_y \delta n\rangle/\langle n_e\rangle$ wave simulations. }
\label{tab:wave_D}
\end{table}

The variation of the normalized anomalous drag term $\langle \delta n_e \delta E_y \rangle/\langle n_e \rangle$ is shown in Table~\ref{tab:wave_D}. Similar to $\delta E_{rms}$, there is an increase of this term with the mass ratio. However, the scaling is stronger, with an order of magnitude increase going from $m_i/m_e = 100$ to $1836$. There is no clear trend in the variation with $\wpwc$, with an increase with $\wpwc$ at mass ratio $400$ and a decrease at realistic mass ratio. While the electric field fluctuations scale like $\sqrt{m_i/m_e}$ as shown earlier, the density fluctuations in the simulations do not change with the mass ratio. 

The additional increase of the drag term with mass ratio can be explained by the phase difference between $\delta n_e$ and $\delta E_y$ \cite{yoo:2024}. We examine the quasi-linear expression for the anomalous collision frequency from \cite{davidson:1975},
\begin{equation}
\nu = \text{Im}\left[ k_y \frac{4\omega_{pi}^2}{k_y^2 v_{thi}^2}\zeta_i Z\left(\zeta_i\right)\right]_{k_y=k_M} \frac{\mathcal{E}}{m_e u_{ey} n},
\end{equation}
where $\zeta_i = \omega/(k v_{thi})$, $u_{ey}$ is the electron velocity driving the instability, $Z$ is the plasma dispersion function and $k_M$ is the wavenumber at maximum growth. Assuming $k_M\rho_e \sim 1$, after some manipulation and normalization, the anomalous drag term is proportional to $(\wpwc)^2/(1+(\wpwc)^2) \sqrt{m_i/m_e}\text{Im}(\zeta_i Z(\zeta_i))$. We find that $\zeta_i < 1$ for all but the $\wpwc = 8$, $m_i/m_e = 1836$ case, and $\zeta_i$ increases with mass ratio, such that $\text{Im}(\zeta_i Z(\zeta_i))$ increases for the studied parameters (the maximum $\text{Im}(\zeta_i Z(\zeta_i))$ is at $\zeta_i \approx 0.7$). 


Although this work focuses on the amplitudes of electrostatic lower-hybrid drift waves in guide-field reconnection, there are wider implications, both for lower-hybrid waves and kinetic simulations in general. Generalizing the setup to 3D allows the excitation of lower-hybrid drift waves with a small $k_\parallel$ with parallel electric fields that accelerate electrons \cite{lavorenti:2021,ng:2023,marshall:2022,cairns:2005}. Variations in wave amplitudes would naturally affect the expected electron energies. We also briefly address the cancellation of the drag term due to anomalous viscosity \cite{graham:2022,price:2020}. In our simulations, the ratio between the magnitude of the drag and viscous terms is on average $\approx 1.3$, such that the cancellation is not perfect. It is likely that this result is due to the shorter wavelengths seen in our simulations ($k\rho_e \approx 1$) compared to the magnetopause observations ($k\rho_e \approx 0.4$) \cite{graham:2022}, meaning that electrons are not as strongly magnetized. Magnetotail observations show agyrotropic distributions where electron gyroradii have similar scales to the wavelength \cite{chen:2020}, suggesting that the waves can contribute to momentum balance there.

In antiparallel or weak guide field reconnection simulations, electrostatic lower-hybrid waves are found outside the current sheet \cite{daughton:2003,mozer:2011}, and do not contribute significantly to the reconnection electric field at the x-line. Nonetheless, the conclusions of this work apply to the wave amplitudes. Eigenmode studies \cite{daughton:2003} begin with a Harris equilibrium \cite{harris:1962}, where the relative velocity between electrons and ions $V_d \propto T/(e B_0 L)$ where $e$ is the unit charge and $L$ is the width of the current sheet. One may perform a similar analysis for the electric field and find that $\delta E/(B_0 v_{A0}) \sim ((\beta_i + \beta_e)/L)(c/\omega_{LH})/\sqrt{4(1+\omega_{pe}^2/\omega_{ce}^2)}$. If $L$ is $d_i$ scale, the mass ratio does not affect the normalized electric field, but if $L$ is $d_e$ scale, the factor of $\sqrt{m_i/m_e}$ reappears. In an evolving system, the current layer thins to sub-ion scales during reconnection, and the electron drift velocity increases, suggesting that the relative effects of the wave electric field would increase when using realistic parameters. Systems with electron-scale reconnection also see stronger flows, though the generation of the LHDI may be limited by their spatial extent. While waves at the separatrix or outside the current sheet may not be important for reconnection, observations show that they  still affect the electron temperature and mixing \cite{wang:2022lhdi,graham:2022,le:2018}.


To summarize, we have shown that even though the reconnection rate in ion-scale regions is not sensitive to the reduced parameters used in kinetic simulation,
the global effects of lower-hybrid drift waves are underestimated, both in terms of their amplitudes relative to the reconnection electric field and the contribution to momentum balance through correlated density and electric field fluctuations. 
The results of this paper are applicable to simulations of  other systems where electron-scale waves and instabilities couple to larger scales. For example, in collisionless shock simulations, fluctuating electric fields are much smaller than the quasi-static electric fields, in contrast to observations where the opposite is true \cite{wilson:2021,bohdan:2024}. Our results suggest that a careful analysis of how waves develop and saturate in multi-scale simulations is necessary, as current simulations could be underestimating their relative effects.

This work was supported by the MMS Mission, NASA Grants 80NSSC21K1462, 80NSSC21K1795, NNH20ZDA001N, and 80NSSC24K0094, and NSF Grant AGS2010231. Simulations were performed using NASA HECC resources.

\bibliography{reconnectionbib}


\end{document}